\documentclass[aip,reprint]{revtex4-1}

\usepackage{amsmath}
\usepackage{amsfonts}
\usepackage{amssymb}
\usepackage{graphicx}
\usepackage{dcolumn}

\newcommand{\ket}[1]{\lvert #1 \rangle}
\newcommand{\bra}[1]{\langle #1 \rvert}
\newcommand{\expval}[3]{\bra{#1}#2\ket{#3}}

\draft 

\begin{document}


\title{Magnitude of pseudopotential localization errors in fixed node diffusion quantum Monte Carlo} 

\author{Jaron T. Krogel}
\email{krogeljt@ornl.gov
\newline\newline
This manuscript has been authored by UT-Battelle, LLC under Contract No. DE-AC05-00OR22725 with the U.S. Department of Energy. The United States Government retains and the publisher, by accepting the article for publication, acknowledges that the United States Government retains a non-exclusive, paid-up, irrevocable, worldwide license to publish or reproduce the published form of this manuscript, or allow others to do so, for United States Government purposes. The Department of Energy will provide public access to these results of federally sponsored research in accordance with the DOE Public Access Plan (http://energy.gov/downloads/doe-public-access-plan).
}
\affiliation{ Materials Science and Technology Division, Oak Ridge National Laboratory, Oak Ridge, Tennessee 37831, USA}
\author{P. R. C. Kent}
\affiliation{ Center for Nanophase Materials Sciences, Oak Ridge National Laboratory, Oak Ridge, Tennessee 37831, USA}
\affiliation{ Computer Science and Mathermatics Division, Oak Ridge National Laboratory, Oak Ridge, Tennessee 37831, USA}

\date{30 March 2017}

\begin{abstract}
Growth in computational resources has lead to the application of real space diffusion quantum Monte Carlo (DMC) to increasingly heavy elements.  Although generally assumed to be small, we find that when using standard techniques the pseudopotential localization error can be large, on the order of an electron volt for an isolated cerium atom.  We formally show that localization error can be reduced to zero with improvements to the Jastrow factor alone and we define a metric of Jastrow sensitivity that may be useful in the design of pseudopotentials.  We employ an extrapolation scheme to extract the bare fixed node energy and estimate the localization error in both the locality approximation  and the T-moves schemes for the Ce atom in charge states $3+$ and $4+$. The locality approximation exhibits the lowest Jastrow sensitivity and generally smaller localization errors than T-moves, although the locality approximation energy approaches the localization free limit from above/below for the $3+/4+$ charge state.  We find that energy minimized Jastrow factors including three-body electron-electron-ion terms are the most effective at reducing localization error for both the locality approximation and T-moves. Less complex or variance minimized Jastrows are generally less effective.  Our results suggest that further improvements to Jastrow factors and trial wavefunction forms will be necessary to reduce localization errors to chemical accuracy in calculations of heavy elements. 
\end{abstract}


\maketitle

\section{Introduction}

Quantum Monte Carlo\cite{foulkes2001} (QMC) is emerging as a successful and accurate first principles approach to ground state problems in both molecular and solid state electronic structure. Recent successful applications include the description of van der Waals binding in molecules\cite{dubecky2016} and the lattice, charge, and spin degrees of freedom in transition metal oxides and selenides\cite{kolorenc2008,kolorenc2010,mitas2010,schiller2015,foyevtsova2014,wagner2014,santana2015,yu2015,zheng2015,wagner2015,benali2016,santana2016,busemeyer2016}.

An appealing feature of the real space diffusion Monte Carlo\cite{grimm1971,anderson1975} (DMC) method, is that it is formally exact: all approximations are subject to systematic improvement, at least in principle.  In practice some approximations can be controlled to a desired level of accuracy today (e.g. finite timestep and population control approximations) while others remain fundamentally more difficult to minimize (e.g. pseudopotential and fixed node\cite{anderson1975,anderson1976} approximations).  For many applications, the error incurred by making these uncontrolled approximations is assumed, and sometimes can be shown, to be small, for example approaching chemical (1 kcal/mol) accuracy for energy differences such as van der Waals binding energies.\cite{dubecky2016}  For general systems, however, the impact of these approximations is not fully known.  Particularly understudied are errors relating to pseudopotential approximations in DMC.

Unlike other wavefunction based approaches to electronic structure (e.g. Hartree-Fock\cite{hartree1928,fock1930,roothaan51}, CISD\cite{pople1977}, CASSCF\cite{roos1980}, CCSD(T)\cite{raghavachari1989,bartlett1990,knowles1993}), the errors due to pseudopotentials in DMC are not limited to those relating to the underlying pseudopotential construction.  Even a pseudopotential that is ``valence perfect'', i.e. one that reproduces all properties of the all electron Hamiltonian, can still produce incorrect energetics in DMC.  The source of these deviations resides in the projection process employed in DMC.  

Diffusion Monte Carlo approaches the ground state wavefunction asymptotically by applying the projection operator $\exp(-t\hat{H})$ to a trial wavefunction $\Psi_T$, where $\hat{H}$ is the Hamiltonian and the importance sampling transformation\cite{grimm1971} has been neglected for simplicity.  When a non-local operator, such as a pseudopotential, is present in $\hat{H}$, the projector can develop negative signs along certain paths in the random walk and the projection operation can no longer be interpreted as a probability density.  This precludes the use of Monte Carlo sampling without further approximation and can be viewed as a second ``sign problem'' in addition to the more widely appreciated fermion sign problem that derives from the antisymmetry requirement of fermionic states.  Practical approximations avoid this non-local sign problem through full or partial localization of the pseudopotential, as in the locality approximation \cite{hurley1987,mitas1991} and the T-moves approach \cite{casula2005,casula2006}, respectively (we refer to both approaches as ``localization'' approximations throughout this work).  Although both approximations become formally exact in the limit of an exact trial wavefunction, in practice the DMC energy gains some dependence on wavefunction details beyond the nodal structure.  Estimating the magnitude of localization error remains difficult, because in standard calculations this error cannot be disentangled from fixed node error. 

In this study, we first review the formal basis for localization error. By considering the fixed node and localization approximations as successive artificial potentials we show that localization errors can be separated from fixed node errors and that they can in principle be eliminated through improvements to the Jastrow factor.  We discuss how localization error can be estimated in practice via an extrapolation approach that is a variant of the one employed in Ref. \onlinecite{casula2005}. Our approach uses both the DMC energies and the energies obtained from the simpler variational Monte Carlo (VMC) approach that does not suffer from locality errors. We then investigate a test problem that shows a large dependence of the DMC energy on variations in the Jastrow factor: the $3+$ and $4+$ charge states of the Ce atom calculated with a moderate valence ($Z_{eff}=12$) pseudopotential.  Using the extrapolation approach we quantify the Jastrow sensitivity and directly estimate the magnitude of localization error with both the locality approximation and the T-moves approach. The strong sensitivities that we find are consistent with a growing understanding that localization errors can no longer be ignored as one descends down the periodic table \cite{nazarov2016,doblhoff2016,drummond2016}.

\section{Overview of localization approximations, extrapolation, and sensitivity}
\label{sec:theory}

In order to explore the sensitivity of the DMC energy on the trial wavefunction, we begin by reviewing some of the basic facts surrounding the fixed node wavefunction.  The results below apply equally well to the fixed phase method, but we follow the fixed node approach for simplicity of discussion. 

We consider the case of a many-body Hamiltonian including a non-local potential
\begin{align}
\hat{H} = \hat{T}+\hat{V}+\hat{V}_{NL}
\end{align}
with fermionic ground state wavefunction and energy denoted $\Psi$ and $E$, respectively.  Here $\hat{T}$ represents the kinetic energy, $\hat{V}$ is a local potential, and $\hat{V}_{NL}$ is a non-local potential arising from the introduction of semi-local pseudopotentials.

The fixed node DMC method solves for the ground state, $\Psi_{FN}$, of a new Hamiltonian that constrains its eigenstates to become exactly zero at and beyond the boundary of the nodal surface of a given antisymmetric trial wavefunction $\Psi_T$
\begin{align}
  &\hat{H}_{FN}\Psi_{FN}=E_{FN}\Psi_{FN}\\
  &\hat{H}_{FN}=\hat{H}+\hat{V}_{FN}(\Psi_T)
\end{align}  
  The fixed node potential, $\hat{V}_{FN}(\Psi_T)$, can be represented as
\begin{align}
  V_{FN}(\Psi_T) =\left\{
    \begin{array}{ll}
      0      & \textrm{when}~ \Psi_T>0           \\
      \infty & \textrm{otherwise}
    \end{array}
    \right.
\end{align}
The fixed node energy is variational $E_{FN}\ge E$, becoming exact only when the nodes of $\Psi_T$ and $\Psi$ coincide.  It is important to note that despite the fact that $\Psi_{FN}$ exists, and can in principle be obtained, difficulties arise due to the non-local potential ($\hat{V}_{NL}$) if one attempts to directly use $\hat{H}_{FN}$ for Monte Carlo projection.

In order to avoid a sign problem in the projector, $e^{-t\hat{H}_{FN}}$, practical DMC calculations involving non-local pseudopotentials make an additional full\cite{hurley1987,mitas1991} or partial\cite{casula2005,casula2006} localization approximation.  If $\hat{V}_{NL}$ represents non-local part of the potential and $\hat{V}_{NL}^{+}$ is the component of $\hat{V}_{NL}$ with positive off diagonal elements in real space (i.e. $\expval{R}{\hat{V}_{NL}}{R'}>0$), then the locality approximation (LA) and the T-moves (TM) approach each introduce an additional modification\cite{bajdich2009} to the fixed node Hamiltonian, yielding
\begin{align}
  \hat{H}_{FN}^{LA} &= \hat{H}_{FN}+\left(\frac{\hat{V}_{NL}\Psi_T}{\Psi_T}-\hat{V}_{NL}\right) \\
  \hat{H}_{FN}^{TM} &= \hat{H}_{FN}+\left(\frac{\hat{V}_{NL}^+\Psi_T}{\Psi_T}-\hat{V}_{NL}^+\right) 
\end{align}
Each localized Hamiltonian bears an explicit dependence on $\Psi_T$ and each localization approximation becomes exact as $\Psi_T$ approaches $\Psi_{FN}$, i.e. if $\Psi_T=\Psi_{FN}$, $\hat{H}_{FN}^{LA}{\Psi_{FN}}=\hat{H}_{FN}^{TM}{\Psi_{FN}}=\hat{H}_{FN}{\Psi_{FN}}=E_{FN}{\Psi_{FN}}$.  As a result we have the following triple equality when $\Psi_{FN}$ is (hypothetically) used as the trial function in QMC calculations
\begin{align}\label{eq:triple_eq}
  E_{VMC}(\Psi_{FN})=E_{DMC}^{LA}(\Psi_{FN})=E_{DMC}^{TM}(\Psi_{FN})
\end{align}
Equation \ref{eq:triple_eq} clearly represents the saturation limit where localization errors have fully been removed but the fixed node error remains untouched.  Approaching this limit is key to assessing the magnitude of localization errors. 

Trial wavefunctions used in most current applications of DMC are of the form $\Psi_T=e^{J_T}\Phi_T$, where $J_T$ is a symmetric Jastrow\cite{jastrow1955} correlation function and $\Phi_T$ contains antisymmetric orbital-based information (such as a single Slater determinant\cite{slater1929} or a multideterminant expansion).  Since $\Psi_T$ and $\Psi_{FN}$ share a nodal surface the ratio $\Psi_{FN}/\Phi_T$ is a symmetric non-negative function, fulfilling the general requirements of a Jastrow factor.  Thus the fixed node wavefunction can be represented in a form reminiscent of the trial wavefunction\cite{holzmann2016}
\begin{align}\label{eq:dmc_jastrow}
  \Psi_{FN} = e^{J_{FN}}\Phi_T
\end{align}
Taking Eqs. \ref{eq:triple_eq} and \ref{eq:dmc_jastrow} together, we see that improvements to the Jastrow factor alone are sufficient to remove localization errors from the DMC energy, at least in principle. 

In practice, useful information about localization error can be obtained by extrapolating VMC and DMC energies over a range of Jastrow factors of increasing sophistication.  We employ a variant of the approach used in Refs. \onlinecite{casula2005} and \onlinecite{casula2006}.  Specifically, we construct a set of Jastrow factors $\{J_n\}$ and perform a series of calculations to obtain sets of QMC energies
\begin{align}
  \{J_n\} \rightarrow & \{E_{VMC}(e^{J_n}\Phi_T)\}, \nonumber\\ 
                      & \{E_{DMC}^{LA}(e^{J_n}\Phi_T)\}, \\ 
                      & \{E_{DMC}^{TM}(e^{J_n}\Phi_T)\} \nonumber
\end{align}
and then arrange the data to allow convenient fitting and extrapolation to the convergence limit of Eq. \ref{eq:triple_eq}. One way of arranging the data for extrapolation is to fit the DMC energies as a function of the VMC energy.  The localization error free value must lie on the $E_{VMC}=E_{DMC}$ line.  Arranging the extrapolants and the VMC/DMC line graphically permits the extraction of three independent estimates of the bare fixed node energy at the intersection points corresponding to $E_{VMC}=E_{DMC}^{LA}$, $E_{VMC}=E_{DMC}^{TM}$, and $E_{DMC}^{LA}=E_{DMC}^{TM}$.    

It is useful to define the ``Jastrow sensitivity'' as the amount of energy decrease observed in the DMC energy per unit decrease in the VMC energy, 
\begin{align}
  S_{J}\equiv \frac{\Delta E_{DMC}}{\Delta E_{VMC}}
\end{align}
This sensitivity can be obtained approximately from a linear fit of the Jastrow series data and it will remain servicable so long as the data do not deviate too strongly from linear dependence.  The ideal situation is for this sensitivity to be small for a given pseudopotential for all relevant valence environments.  If this is the case, the Jastrow factors and wavefunction optimization methods available today are more likely to reduce localization errors to an acceptable level. This may be particularly important when comparing energies between systems with very different electronic structures, such as low and high pressure phases of a material. Given the large sensitivity and residual localization error observed in this study, we anticipate that Jastrow sensitivity defined in this way will provide a useful metric for pseudopotential design efforts that specifically target DMC.

\begin{figure*}[t]
  \includegraphics[scale=0.5,trim={1.0in 0.9in 0 0},clip]{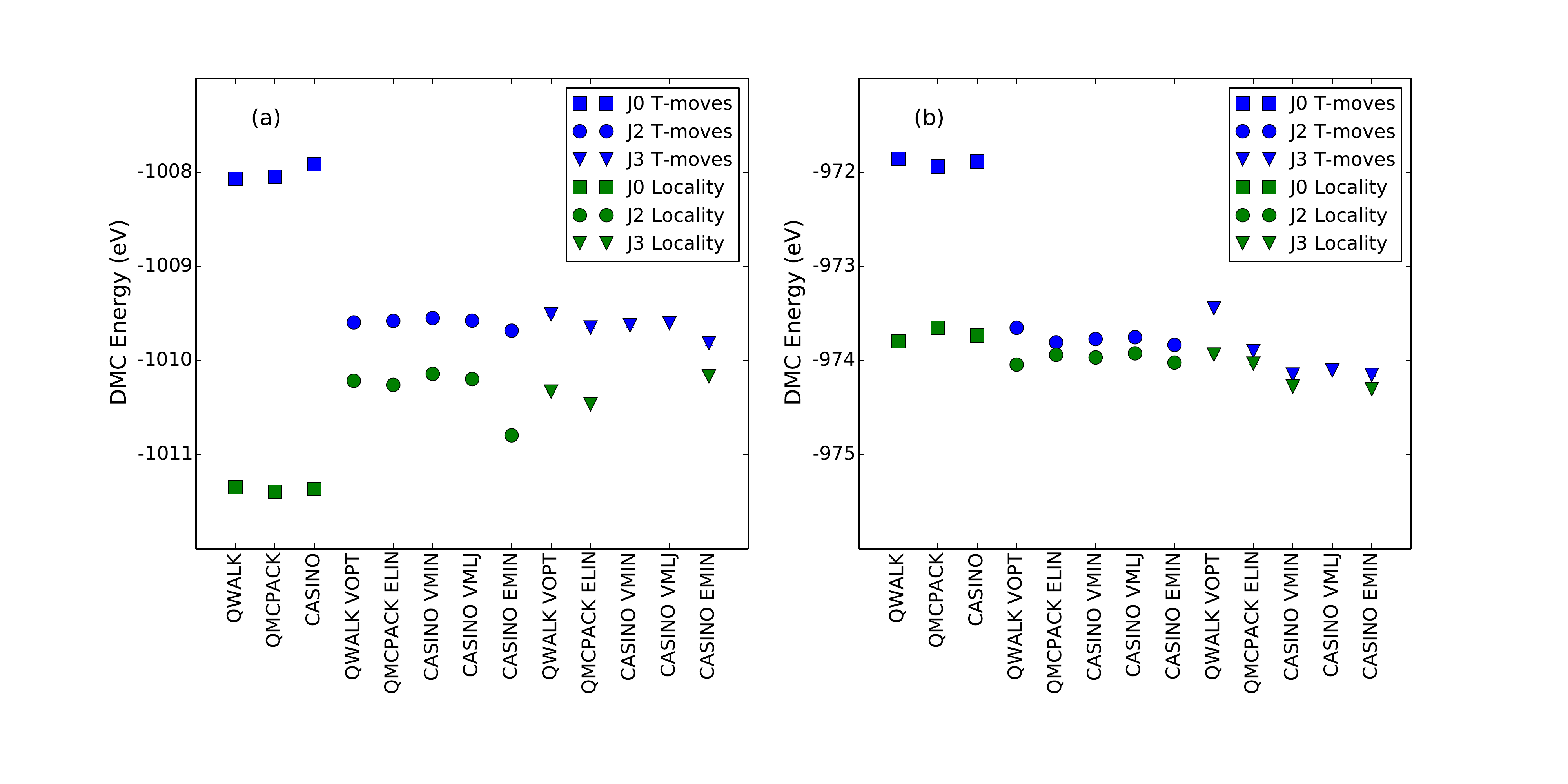}
  \caption{DMC total energies for the Ce pseudo-atom in $3+$ (a) and $4+$ (b) charge states (valence configurations $[Pd]5s^25p^64f^1$ and $[Pd]5s^25p^6$) for various Jastrow factors optimized with open QMC codes (``$J_0$'' means no Jastrow).  In the absence of localization error, all points would fall on a horizontal line.}
  \label{fig:qmc_code}
\end{figure*}

\section{Calculation details}
\label{sec:calc}

Diffusion Monte Carlo calculations were performed for an isolated cerium atom in two positively charged states: the $3+$ state with valence configuration $[Pd]5s^25p^64f^1$ and the $4+$ state with valence configuration  $[Pd]5s^25p^6$.  The Ce pseudopotential employed here\cite{dolg_cepp} is of Dirac-Fock type with a Pd core (12 valence electrons)\footnote{This pseudopotential was also used in a recent LRDMC study\cite{devaux2015} of metallic cerium.  LRDMC\cite{casula2005} differs in some details from standard DMC and the large localization errors we observe (standard DMC) may have been further mitigated in that study through the use of advanced geminal wavefunctions.}. In the DMC calculations $f$ was used as the local channel.

Single particle orbitals were obtained from GAMESS\cite{schmidt93,gordon05} or CRYSTAL\cite{dovesi2014} DFT calculations with the PBE0\cite{adamo1999} functional.  DFT total energies between the two codes agreed to better than 0.05 eV for both charge states.  In the GAMESS calculations, we used the Ce basis set of Graciani \emph{et al.}\cite{graciani2011} with the $g$ function removed, resulting in an uncontracted basis of size $10s10p7d8f$.  For the CRYSTAL calculations the basis set size was reduced to $7s6p5d8f$ by removing the most core-like basis functions (the original basis set was designed for a Ce pseudopotential with $Z_{eff}=30$).  Hartree-Fock calculations within GAMESS showed that the effect of removing the core basis functions was on the order of 0.05 eV.  We also explored basis sets up to size $21s23p15d16f6g$ in GAMESS and basis set optimization with CRYSTAL\footnote{The basis set optimization was carried out with the ``billy'' wrapper script for CRYSTAL that is distributed with the CASINO QMC code.} and found very small improvements ($\sim$ 0.03 eV) over the basis sets used here.

A series of quantum Monte Carlo calculations were performed with three openly available QMC codes: QMCPACK\cite{kim2012}, CASINO\cite{needs2010}, and QWALK\cite{wagner2009}.  Both two- and three-body Jastrows were optimized with a range of optimizers available in the three codes including standard variance minimizers\cite{umrigar1988}, a global minimum log-linear Jastrow variance minimizer\cite{drummond2005}, and implementations of the linearized optimization method\cite{umrigar2007} for energy and variance.  QMCPACK optimizations were performed with the linear method with an 80/20 mix of energy and variance in the cost function (denoted ELIN here).  The OPTIMIZE method (denoted VOPT) was used in QWALK to perform variance minimization.  QWALK's OPTIMIZE2 method was also employed, but the resulting VMC energies were higher than those obtained without a Jastrow.  Three methods were used with CASINO: VARMIN/VARMINLINJAS (denoted VMIN/VMLJ) for variance minimization and EMIN for pure energy minimization.  Diffusion Monte Carlo calculations were performed with 2048 walkers and a projector discretization timestep of $0.0025~Ha^{-1}$, resulting in an acceptance ratio of better than $99.8\%$ in all cases.  As we will see below, this approach gives results that can be grouped by degree of sophistication of the Jastrow factor, with smaller variations within each set due to details of the specific parameters and optimization used.

\section{Assessing localization errors for the Ce atom}
\label{sec:results}

The results of our QMC calculations of atomic Ce are summarized in Fig. \ref{fig:qmc_code}.  In Fig. \ref{fig:qmc_code}, DMC total energies within T-moves (blue) and the locality approximation (green) are shown for a set Jastrow factors of varying quality obtained with QMCPACK, CASINO, and QWALK.  The method of optimization is indicated by the labels.  Missing points for the locality approximation correspond to CASINO runs with unstable (diverging) walker populations.

Localization errors are quite significant for both the $3+$ and $4+$ charge states of the pseudo-atom.  Generally, if localization errors were negligible, all data points would reside at the same value regardless of the Jastrow factor or the employed localization constraint.  The DMC energies vary on the scale of tenths of an eV, which presents a significant challenge if one is aiming for chemical accuracy (1 kcal/mol$\approx$ 0.04 eV), as is often the case in real world applications.  The sizable variations seen here indicate that localization errors are large, but they also show that these errors can have a strong dependence on the valence configuration of the atom and this has direct implications for larger scale calculations of molecules or solids. 

For the $3+$ charge state (see Fig. \ref{fig:qmc_code}a), the difference in DMC total energies between T-moves and the locality approximation is over 3 eV when no Jastrow factor is used.  This difference diminishes to less than 1 eV for optimized Jastrow factors, but it generally remains above 0.6 eV.  For partial localization (T-moves), optimizations involving the variance perform similarly well, while pure energy minimization consistently yields the lowest DMC energy.  For full localization, it appears that better Jastrows (in the T-moves sense) typically lead to higher energies.  This is, of course, consistent with the fact that the locality approximation is non-variational.  Adding a three-body Jastrow does not lead to much improvement, which is somewhat surprising since three-body Jastrows can contribute structure that is missing from the typical isotropic two-body and ion-centered s-wave one-body terms.

Upon removal of the $4f$ electron ($4+$ state) the picture changes most for the locality approximation (see Fig. \ref{fig:qmc_code}b).  There is still a large difference (~2 eV) between T-moves and locality approximation energies without a Jastrow.  When optimized Jastrow factors are used, the spread between TM/LA energies dimishes substantially, often remaining within 0.2 eV.  It may be tempting to conclude that localization errors are relatively small for both TM and LA in this case, but this is misleading.  Due to the non-variational nature of the locality approximation, $E_{DMC}^{TM}-E_{DMC}^{LA}$ may over- or underestimate the actual magnitude of the localization error.  It is clear, however, that the locality approximation results are more weakly dependent on the Jastrow factor for the  $4+$ state than for the $3+$ state, so it is reasonable to assume that the residual errors are also smaller in this case.  In addition to displaying a weak Jastrow dependence, the LA energies now exhibit a roughly decreasing trend.  This apparent switch in trend behavior raises concerns for error cancellation within the locality approximation if energy differences are taken across systems that vary in electronic structure. 

Overall the sensitivity to variations in the Jastrow factor for this system is worrisome, and it is possible that that localization errors may dominate over fixed node errors.  This is important in general since localization errors have long been assumed to be small\cite{mitas1991,foulkes2001}.  We next try to isolate localization errors from fixed node errors via extrapolation.

\begin{figure*}
  \includegraphics[scale=0.5,trim={1.0in 0 0 0 },clip]{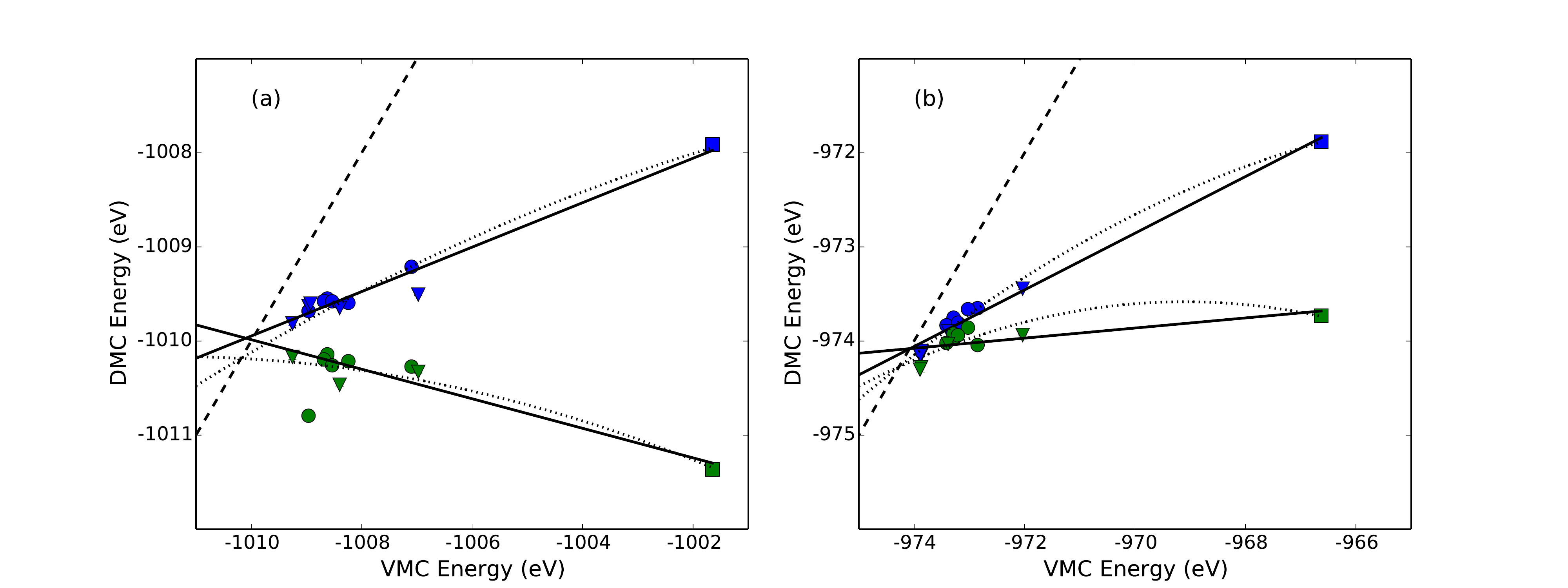}
  \caption{Linear extrapolation of QMC energies toward the localization error free limit for the Ce pseudo-atom in $3+$ (a) and $4+$ (b) charge states (valence configurations $[Pd]5s^25p^64f^1$ and $[Pd]5s^25p^6$ ).  Solid lines are independent linear least squares fits to QMC data and the dashed line represents $E_{VMC}=E_{DMC}$.  Dotted lines show a biquadratic fit (see Eq. \ref{eq:biquad}).  As in Fig. \ref{fig:qmc_code}, TM/LA data are shown in blue/green with the Jastrow factor form indicated by the symbol shape ($J_0$: square, $J_2$: circle, $J_3$: triangle). }
  \label{fig:qmc_extrap}
\end{figure*}

In Fig. \ref{fig:qmc_extrap} the data is rearranged as described in Sec. \ref{sec:theory} with DMC energies plotted vs. their VMC counterparts.  This arrangement more clearly displays patterns in the data. As expected, DMC TM and LA energies generally approach each other as the VMC energy decreases.  On the plots, the condition $E_{DMC}=E_{VMC}$ is represented as a dashed line. This would be obtained for a perfect Jastrow factor.  Solid lines are linear least squares fits to the data and only a single value for the $J_0$ data has been included in the fits (from QMCPACK).  The CASINO $J_2$ EMIN point for Ce $3+$ DMC-LA was excluded from the fit as an outlier, though it serves as an important demonstration of the poorly understood differences between TM and LA: this Jastrow is clearly quite good for TM, but it is poor for LA.

We extract the Jastrow sensitivity $S_J\equiv \Delta E_{DMC}/E_{VMC}$ from the slope of each fit line.  For Ce $3+$ the TM sensitivity is approximately $S^{TM,3+}_{J}=0.24$, meaning that for every 1 eV gained in binding energy during wavefunction optimization the DMC-TM energy is expected to fall by about a quarter of an eV.  Consistent with our observations earlier, the LA sensitivity is opposite in sign:  $S^{LA,3+}_{J}=-0.16$.  For Ce $4+$, the TM sensitivity actually increases $S^{TM,4+}_{J}=0.30$ while the LA sensitivity changes sign and markedly decreases in magnitude $S^{LA,4+}_{J}=0.05$.  Overall, the locality approximation is less sensitive to changes in the Jastrow factor, and therefore yields more accurate total energies, on average, than T-moves for the system studied here.  The difference in sensitivities is much smaller for T-moves ($S^{TM,4+}_{J}-S^{TM,3+}_{J}=0.06$) than it is for the locality approximation ($S^{LA,4+}_{J}-S^{LA,3+}_{J}=0.21$), indicating that T-moves might have better error cancellation--and hence be more accurate--for an energy difference such as the fourth ionization potential of Ce.  It would be particularly interesting to see whether or not a broader study of TM and LA sensitivities for a range of pseudopotentials shows the same pattern for the accuracy of the respective approximations for total and relative energies.  

The triangle of intersection represents the condition in Eq. \ref{eq:triple_eq}, giving three estimates of the fixed node energy with localization errors removed.  From $E_{DMC}^{TM}=E_{VMC}$, $E_{DMC}^{LA}=E_{VMC}$, $E_{DMC}^{TM}=E_{DMC}^{LA}$ we obtain fixed node energies of -1009.93,-1009.99, and -1010.10 eV, respectively, for Ce $3+$.  For Ce $4+$ the corresponding estimates of the fixed node energy are -974.08, -974.08, and -974.07 eV.  Given the shallow slope of the linear fits, it is to be expected that the estimate arising from the condition $E_{DMC}^{TM}=E_{DMC}^{LA}$ has the largest systematic error.  The systematic errors present in these fits can be reduced by performing a single simultaneous fit to TM and LA data that directly satisfies the condition in Eq. \ref{eq:triple_eq}, including terms up to quadratic order to account for slight deviations from linearity observed in the data.  The biquadratic form for the fit is 
\begin{align} \label{eq:biquad}
  E^{TM}_{FN} &= a_{TM}\Delta E_{VMC}^2+b_{TM}\Delta E_{VMC}+E_{FN} \\
  E^{LA}_{FN} &= a_{LA}\Delta E_{VMC}^2+b_{LA}\Delta E_{VMC}+E_{FN} \nonumber
\end{align}
with $\Delta E_{VMC}\equiv E_{VMC}-E_{FN}$ and $a_{TM}$, $b_{TM}$, $a_{LA}$, $b_{LA}$, and $E_{FN}$ as fitting parameters.  A least squares fit to this biquadratic form, shown as dotted lines in Fig. \ref{fig:qmc_extrap},  yields our best estimate of the fixed node energies: $E^{3+}_{FN}\approx -1010.18$ eV and $E^{4+}_{FN}\approx -974.27$ eV.

\begin{table*}
  \addtolength{\tabcolsep}{1mm}
  \begin{ruledtabular}
  \begin{tabular}{@{}l@{}l@{}l@{}r@{}r@{}r@{}r@{}r@{}r@{}r@{}r@{}r@{}r@{}r@{}r@{}}
              &       &     & $\Delta E_{VMC}^{3+}$ & $\Delta E_{VMC}^{4+}$ & $\Delta E_{DMC}^{TM,3+}$ & $\Delta E_{DMC}^{TM,4+}$  & $\Delta E_{DMC}^{LA,3+}$ & $\Delta E_{DMC}^{LA,4+}$ & $\Delta IP^{TM}_{DMC}$ &  $\Delta IP^{LA}_{DMC}$ \\
       QWALK  &       & J0  & 8.37(3)  & 7.64(3)  & 2.11(2)  & 2.39(3)  & -1.16(2)  &  0.54(3)  &  0.12(6)  &  1.72(6)  \\
     QMCPACK  &       & J0  & 8.51(3)  & 7.59(3)  & 2.14(3)  & 2.42(3)  & -1.21(5)  &  0.48(2)  &  0.30(3)  &  1.64(2)  \\
      CASINO  &       & J0  & 8.53(3)  & 7.62(3)  & 2.27(5)  & 2.34(3)  & -1.18(5)  &  0.62(5)  &  0.20(4)  &  1.83(8)  \\
              &       &     &          &          &          &          &           &           &           &           \\
       QWALK  & VOPT  & J2  & 1.94(2)  & 1.42(1)  & 0.59(1)  & 0.62(1)  & -0.03(1)  &  0.23(1)  &  0.03(1)  &  0.26(1)  \\
     QMCPACK  & ELIN  & J2  & 1.65(3)  & 1.07(1)  & 0.61(1)  & 0.47(1)  & -0.07(2)  &  0.33(1)  & -0.14(2)  &  0.41(2)  \\
      CASINO  & VMIN  & J2  & 1.56(1)  & 1.01(1)  & 0.64(2)  & 0.50(2)  &  0.04(5)  &  0.30(2)  & -0.13(3)  &  0.26(6)  \\
      CASINO  & VMLJ  & J2  & 1.50(2)  & 0.99(1)  & 0.61(2)  & 0.52(2)  & -0.01(5)  &  0.35(3)  & -0.09(3)  &  0.36(6)  \\
      CASINO  & EMIN  & J2  & 1.22(2)  & 0.86(1)  & 0.50(2)  & 0.44(2)  & -0.61(5)  &  0.25(2)  & -0.06(3)  &  0.86(6)  \\
              &       &     &          &          &          &          &           &           &           &           \\
       QWALK  & VOPT  & J3  & 3.21(2)  & 2.24(1)  & 0.68(1)  & 0.83(1)  & -0.14(1)  &  0.34(1)  &  0.15(1)  &  0.48(1)  \\
     QMCPACK  & ELIN  & J3  & 1.79(5)  & 0.89(1)  & 0.54(1)  & 0.38(1)  & -0.28(2)  &  0.24(1)  & -0.16(1)  &  0.52(2)  \\
      CASINO  & VMIN  & J3  & 1.22(1)  & 0.40(1)  & 0.56(2)  & 0.13(1)  &           & -0.00(5)  & -0.43(2)  &           \\
      CASINO  & VMLJ  & J3  & 1.26(1)  & 0.41(1)  & 0.58(2)  & 0.17(1)  &           &           & -0.41(2)  &           \\
      CASINO  & EMIN  & J3  & 0.93(2)  & 0.38(1)  & 0.37(2)  & 0.12(1)  &  0.02(3)  & -0.03(3)  & -0.25(3)  & -0.05(4)  \\
  \end{tabular}
  \caption{Estimated VMC errors and DMC localization errors (eV) in total energies and the fourth ionization potential of pseudo-Ce. $\Delta E\equiv E-E_{FN}$ and $\Delta IP\equiv (E^{4+}-E^{3+})-(E^{4+}_{FN}-E^{3+}_{FN})$, with $E_{FN}$ estimated from the biquadratic fit of the VMC and DMC energies. Details of the optimization methods and codes are given in Sec. \ref{sec:calc}.}
  \label{tab:loc_err}
  \end{ruledtabular}
\end{table*}

These extrapolated values have been used to obtain estimates of the localization error for T-moves and the locality approximation for pseudo-Ce as summarized in Table \ref{tab:loc_err}.  The quality of the results is consistent at the two-body Jastrow level.  The estimated localization error for T-moves remains near 0.6 eV for Ce $3+$ and 0.5 eV for Ce $4+$, leading to a 4th ionization potential that deviates by about 0.1 eV from the extrapolated value.  The error in the locality approximation is smaller in absolute terms-- about -0.05 eV for Ce $3+$ and 0.3 eV for Ce $4+$-- but suffers from worse cancellation, yielding an error of about 0.35 eV in the IP.  Including three body terms in the Jastrow factor generally reduces absolute localization errors, but with more spread, sometimes leading to poor error cancellation.  A few of the IP estimates using J3 contain errors on the order of 0.4 eV, which may reflect the greater challenge of optimizing Jastrow forms with larger variational freedom.  

Error cancellation cannot be totally relied on for either level of description for Jastrow factors.  The best overall Jastrow factor was obtained via energy minimization (EMIN J3) and led to the smallest absolute localization errors: 0.37(2)/0.12(1) eV for Ce $3+/4+$ with T-moves and 0.02(3)/-0.03(3) eV for Ce $3+/4+$ with the locality approximation.  The residual error in the IP does not cancel for T-moves (-0.25(3) eV), while it is small for the locality approximation (-0.05(4) eV) purely on the basis that absolute errors are small.  The lack of cancellation on the part of T-moves, despite exhibiting similar sensitivity across charge states, can be traced to the fact that the Jastrow factor obtained for Ce $4+$ was more accurate than the one obtained for Ce $3+$, in the energetic sense (VMC error of 0.93(2) eV for Ce $3+$ vs. 0.38(1) eV for Ce $4+$).  Sufficient error cancellation might be regained in applications where energy differences are taken between systems with similar electronic structure and variations in the Jastrow factor are restrained (this is in fact often exploited in the context of van der Waals binding calculations\cite{DubeckyChemRev2016}).  Although our tests are performed for atoms, we expect the sensitivies to persist in molecular and solid-state systems. Based on our overall results, we recommend that more effort be given to develop improved Jastrow forms to systematically reduce localization errors from DMC calculations of heavy elements.

\section{Summary}

We have explored non-local pseudopotential localization error in DMC at a theoretical and practical level.  We have shown that localization error, in both the locality approximation and the T-moves scheme can be formally removed by approaching the exact Jastrow factor in the presence of the fixed node approximation.  We have also proposed a numerical measure of Jastrow sensitivity that represents the intrinsic challenge posed by a given pseudopotential to reduce localization errors to an acceptable level. Jastrow sensitivity estimates of existing pseudopotentials could serve as a first step to identify where DMC can be safely applied in a routine fashion and highlight where greater effort is needed.  We anticipate it could also be useful to minimize the Jastrow sensitivity as a direct objective during pseudopotential construction to reduce localization errors generally.
 
Using an extrapolation approach we have estimated the Jastrow sensitivity and the magnitude of localization error in the case of a Ce atom in its $3+$ and $4+$ charge states.  We find that wavefunction optimization utilizing energy minimization is an effective means at reducing localization error, with the locality approximation approaching chemical accuracy for the best Jastrow factors.  For these same Jastrow factors we estimate residual localization errors on the order of 3-8 kcal/mol with T-moves.  Consistent with this, we found that DMC energies changed by 0.25-0.30 eV with T-moves and 0.05-0.16 eV with the locality approximation for every 1 eV gained at the VMC level.  In general, the development of better Jastrow factors may be necessary to routinely obtain chemical accuracy with a given localization approach. A more comprehensive study of the relative performance of the locality approximation and the T-moves scheme is warranted and could be effectively carried out following the approach used here.

\section*{Supplementary Material}
See supplementary material for VMC and DMC total energies of the Ce atom in $3+$ and $4+$ charge states.

\section*{Acknowledgements}
This research was sponsored by the Laboratory Directed Research and Development Program of Oak Ridge National Laboratory, managed by UT-Battelle, LLC, for the U.S. Department of Energy. This research used resources of the Oak Ridge Leadership Computing Facility at the Oak Ridge National Laboratory, which is supported by the Office of Science of the U.S. Department of Energy under Contract No. DE-AC05-00OR22725.

\bibliography{ref}

\end{document}